\documentclass[prl,twocolumn,balancelastpage, superscriptaddress, shownopacs]{revtex4}

\usepackage{amsmath,amssymb,amsfonts}
\usepackage{bm,graphicx,color}

\newcommand{\TR}{\text{Tr}}

\newcommand{\Vk}{{\bm k}}

\newcommand{\VA}{{\bm A}}

\newcommand{\CC}{\mathcal{C}}

\newcommand{\hopp}{v}

\usepackage{float}

\begin{document}

  \title{Photo-induced states in a Mott insulator}

\author{Martin Eckstein}
\affiliation{Max Planck Research Department for Structural Dynamics, University of Hamburg-CFEL, Hamburg, Germany}
\author{Philipp Werner}
\affiliation{Department of Physics, University of Fribourg, 1700 Fribourg, Switzerland}
 
  \date{\today}

\begin{abstract}
We investigate the properties of the metallic state obtained by photo-doping carriers into a Mott insulator. 
In a strongly interacting system, these carriers have a long life-time, so that they can dissipate their kinetic energy to a phonon bath. 
In the relaxed state, the scattering rate saturates at a non-zero temperature-independent value, and the momentum-resolved spectral function 
features broad bands which differ from the well-defined quasi-particle bands of a chemically doped system.  
Our results indicate that a photo-doped Mott insulator behaves as a bad metal, in which strong scattering 
between doublons and holes inhibits Fermi-liquid behavior down to low temperature.
\end{abstract}


\maketitle

Carrier doping provides a convenient way of controlling the properties
of strongly correlated materials. The high-T$_\text{c}$ cuprates, e.g., 
undergo a doping-driven transition from an antiferromagnetic Mott insulator to a 
correlated metal that becomes superconducting at low temperatures.
While doping is usually achieved by modifying the chemical composition of a 
material, photo-doping, i.e., a change of the electron and hole concentrations by irradiating light on a sample,
provides a straightforward way to influence material properties in an 
ultra-fast manner. Photo-induced insulator-to-metal transitions have been 
demonstrated in organic materials \cite{Okamoto2007,Wall2011}, charge-transfer 
insulators \cite{Iwai2003} and cuprates \cite{Okamoto2010}.

In contrast to chemically doped states, photo-doped states simultaneously have electron 
and hole-like carriers, and these carriers are typically inserted with large kinetic energy. 
In a semiconductor, this mainly results in a different occupation of valence and conduction 
bands, which themselves are rigid, i.e., independent of filling and temperature. In contrast,  
the formation of a Fermi liquid in a doped correlated system leads to a narrow band of delocalized 
quasi-particle states without any counterpart in the insulator. Because these quasi-particles 
are strongly damped at high temperature, and because the adiabatic correspondence between bare 
electrons and quasiparticle excitations of a Fermi liquid is anyway difficult to reconcile 
with the presence of both holes and electrons, the properties of a correlated photo-doped 
state are not obvious. Its large kinetic energy works against correlations and could cause
a rigid-band behavior. On the other hand, life-times of a few picoseconds indicate that 
carriers are strongly coupled to spin fluctuations or phonons \cite{Iwai2003,Okamoto2010}.
They might thus dissipate their energy before recombination, and reveal correlation effects 
in a photo-doped state and possible differences to a Fermi liquid.

In this paper we contrast photo-doped and chemically doped Mott insulators in the simplest 
possible setup, which is a paramagnetic single-band Mott insulator that is initially perturbed 
by an intense laser pulse. For this purpose we focus on the Hubbard model,
\begin{equation}
\label{hubbard}
H= \sum_{
\langle ij
\rangle
,\sigma=\uparrow,\downarrow} \!\!
\hopp_{ij}
\, c_{i\sigma}^\dagger c_{j\sigma}
+ U
\sum_{i}
n_{i\uparrow}
n_{i\downarrow}
-
\mu 
\sum_{i\sigma}
n_{i\sigma},
\end{equation}
which describes electrons that hop between nearest neighbors $\langle ij\rangle$  of a crystal lattice 
and interact through a local Coulomb repulsion $U$. 
The electric field of the laser pump is determined by the vector potential, $\bm{E}=-\frac{1}{c}\partial_t \bm{A}$,
which in turn enters
Eq.~(\ref{hubbard}) by the Peierls substitution, 
i.e., band energies $\epsilon_\Vk$ are shifted to $\tilde\epsilon_\Vk= \epsilon_{\Vk-ea/\hbar 
c \VA(t)}$ (see, e.g., Ref.~\cite{Davies1988}).
If the Mott gap is large ($U\gg \hopp$), electron-hole pairs have a long lifetime, 
because the electronic decay-channel involving a direct transfer of interaction energy into 
kinetic energy becomes inefficient \cite{Sensarma2010a}.
To account for dissipation of energy to other degrees of freedom, we 
weakly couple our model to a bosonic bath.
As we will show below, the resulting
reduction of the kinetic energy of the particles 
drives the system into a metallic state that is not accessible in thermal equilibrium. 

The model is treated within the nonequilibrium extension \cite{Schmidt2002,Freericks2006} of 
dynamical mean-field theory  (DMFT) \cite{Georges1996}. A direct way to incorporate dissipation 
into this formalism is to attach one particle-reservoir to each lattice site \cite{Tsuji2009,Amaricci2011,
Werner2012}. Here we instead use baths of harmonic oscillators, such that 
 the particle number remains constant while energy is exchanged with the environment. The electronic self-energy, which is 
a functional of the local Green function  $G$ in DMFT, is then the 
sum of the electronic contribution $\Sigma_\text{U}[G]$, and a bath contribution. 
Vertex corrections are neglected for weak coupling to the bath.
The bath contribution is the lowest order diagram for a Holstein-type electron-phonon coupling,  $\Sigma_\text{diss}[G]=\lambda^2 
G(t,t') D(t,t')$, where
$\lambda$ measures the coupling strength, and
 $D(t,t')$ is the equilibrium propagator for a boson with energy $\omega_0$ at given 
temperature $1/\beta$ (such that the bath has no memory); following  the notation of Ref.~\cite{Eckstein2010} 
for contour-ordered Keldysh Green functions, $D(t,t')=-i \TR [ T_\CC \exp(-i \int_\CC dt \omega_0 b^\dagger b) 
b(t)b^\dagger(t')]/Z $ ($\omega_0=\hopp$ in the following). 

Local diagrams for $\Sigma_\text{U}[G]$ are summed to all orders by solving an auxiliary 
single-impurity Anderson problem, just as it is done for DMFT without dissipation.
Below we will use the self-consistent strong-coupling expansion \cite{Eckstein2010b}
to solve this impurity model. Because dissipation introduces a timescale $\propto 1/\lambda^2$ 
to the dynamics, which may be much longer than the hopping time (and which we must access 
within our simulation), we are restricted to the lowest order of the strong-coupling expansion, 
the non-crossing approximation (NCA) \cite{NCA}. However, our main results are obtained 
for a (doped) Mott insulator at $U\gg \hopp$, above and close to the Fermi-liquid coherence 
temperature, where the use of NCA is still justified
($U=14\hopp$, as for the organic salt $\rm{ ET}$-$\rm{F_2TCNQ}$). 
The resulting  equations have been explained in Ref.~\cite{Eckstein2011}. 
To include dissipation as described above we just have to replace the band energy 
$\tilde \epsilon_{\Vk}$ by $\tilde \epsilon_{\Vk} + \Sigma_\text{diss}[G]$ in those equations.

DMFT is exact for infinite dimensions \cite{Metzner89} and yields the generic behavior of a 
high-dimensional lattice, independent of the lattice geometry used within the calculation (mainly, 
the energy scales are renormalized for different geometries). In this paper we perform all calculations 
for the paramagnetic phase of a one-dimensional system, because this drastically simplifies the 
momentum summations, but we expect the results to be representative of high-dimensional systems. 
The unit of energy is set by the hopping $\hopp$, and time is measured in units of $\hbar/\hopp$. In these 
units, the critical end-point of the first-order Mott transition line for the half-filled system is located 
at $U \approx 4$-$5$. 
The pump pulse, $E(t)=E_0 \sin[\Omega(t-t_0)] \exp[-4.605\times(t-t_0)^2/t_0^2]$, is taken to be 
a  sine-wave with frequency $\Omega$, amplitude $E_0$, and Gaussian envelope. The number
of cycles, $N=t_0\Omega/\pi$, is taken between 2 and 10 below. To analyze
photo-doped and equilibrium states, we compute the real-time optical conductivity 
\cite{Eckstein2008} and its partial Fourier transform,
$\sigma(\omega,t)
=
\int_0^{s_\text{max}}
\!ds \,\,
\sigma(t,t-s) e^{i\omega s}$.
We also define an average of $\sigma(\omega,t)$ which is more local in time and frequency
(and almost independent of the cutoff $s_\text{max}=t$),
\begin{equation}
\bar \sigma(\omega,t;\Delta)
\equiv
\int_0^{s_\text{max}}
\!ds \,\,
e^{-s^2 \Delta^2/2}
\sigma(t,t-s) e^{i\omega s},
\label{sigma-av}
\end{equation}
equivalent to broadening $\sigma(\omega,t)$ with  $\exp({-\omega^2 /2 \Delta^2 })$. 

\begin{figure}[t]
\centerline{ \includegraphics[clip=true,width=0.99\columnwidth]{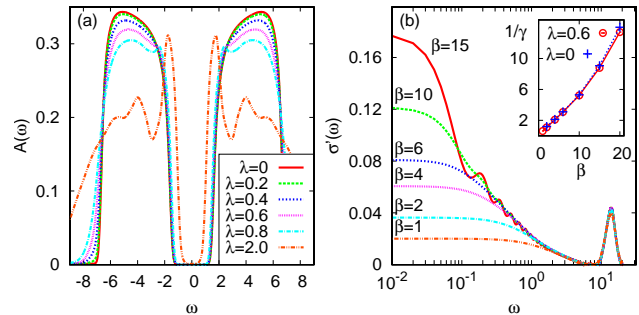}}
\caption{
(a)  Equilibrium spectral function of the half-filled insulator  for $U=8$, $\beta=10$, and various 
$\lambda$. (b)   Frequency-dependent conductivity $\sigma(\omega)$ of the 
doped Mott insulator at various temperatures, $U=14$, $n=1.02$. 
Fourier artifacts due to a finite cutoff $s_\text{max}=40$ appear when the 
peak becomes too narrow ($\beta =15$). The inset shows the Drude relaxation 
time $1/\gamma$, obtained from a Lorentzian fit to $\sigma(\omega)$ for $0<\omega<0.2$.}
\label{fig-1}
\end{figure}  	

Before studying the photo-doped case, we investigate the influence of the dissipative bath on 
equilibrium properties of the (doped) Mott-insulator. Pronounced bath-induced artifacts begin to
appear for $\lambda\gtrsim0.8$, as exemplified for the spectral function of the insulating 
phase in Fig.~\ref{fig-1}a. Because our approximation is only suitable for weak coupling between 
electrons and bath, we will from now on restrict $\lambda$ to values where those features are 
absent and dissipation implies only a slight broadening of the spectrum ($\lambda\lesssim0.6$).
Small electron doping ($n>1$) of the Mott insulator leads to the formation of a narrow quasi-particle 
peak at the lower edge of the upper Hubbard band, which goes along with the emergence of a Drude 
peak in the low-frequency part of the optical conductivity  (Fig. \ref{fig-1}b),
$\sigma(\omega) = D\gamma\pi^{-1}[\gamma^2+\omega^2]^{-1} + \sigma_\text{incoh}(\omega).$
For a Fermi liquid, the scattering rate $\gamma$ decreases with decreasing temperature, with the 
asymptotic behavior $\gamma \sim T^2$. Although the NCA approximation is known to suffer from 
non-causal artifacts deep in the Fermi liquid regime, the onset of this Fermi liquid regime 
can still be observed around $\beta=20$, by extracting $\gamma$ from a fit of $\sigma(\omega)$ with 
a Lorentz curve (Fig.~\ref{fig-1}, inset). For $\beta\lesssim 20$, coupling to a bath with $\lambda=0.6$ 
has only a small influence on the scattering rate.

\begin{figure}[t]
\centerline{ \includegraphics[clip=true,width=1.05\columnwidth]{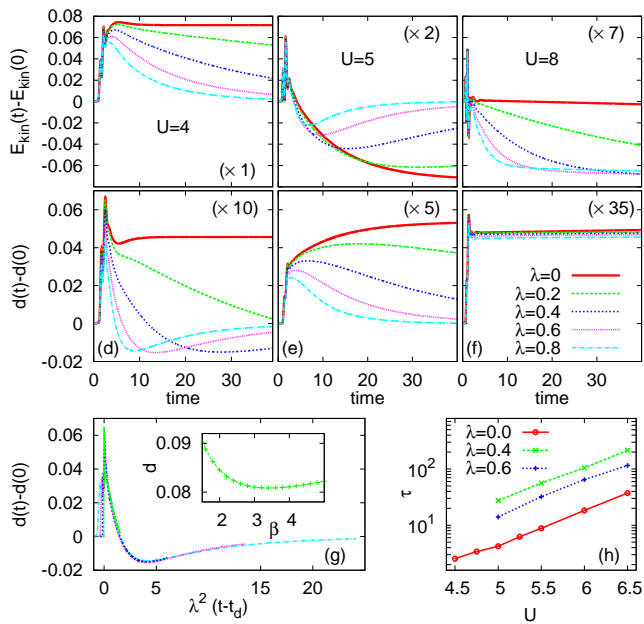}}
\caption{
Relaxation of kinetic energy (a-c) and double occupancy (d-f) for $U=4$ (left), $U=5$ (middle), 
and $U=8$ (right), during and after a pump with frequency $\Omega=U$, $N=10$ cycles, 
$E_0=1$; $\beta=10$. Curves show the difference between the time-evolved values and the initial 
state values, brought to the same scale by an additional factor  (``$\times$'').
Bold solid lines correspond to no dissipation.  
(g) Same data as (d), plotted against rescaled time $t'=\lambda^2(t-t_d)$, where $t_d$ is 
the duration of the pump. The inset shows the equilibrium value of $d$ as a function
of $\beta$ for $U=4$. (h) Relaxation time of the double occupancy, from an exponential fit.}
\label{fig-2}
\end{figure}

In the following paragraphs we study the relaxation of the double occupancy and kinetic energy
(Fig.~\ref{fig-2}a-\ref{fig-2}f) after a pump at $\Omega=U$ has created a small number of charge 
excitations in the system. Without dissipation ($\lambda=0$) and for small $U$ ($U=4,5$), both 
$d(t)$ and $E_\text{kin}(t)$ follow an exponential relaxation $d(t) \sim d_\infty + Ae^{-t/\tau}$ towards 
a value $d_\infty\neq d(0)$.  This fact has already been described for the hyper-cubic lattice in 
Ref.~\cite{Eckstein2011}, where it was also verified that $d_\infty$ is consistent with a thermalization 
of the system at a temperature that is determined by the total amount of absorbed energy. The 
thermalization time exponentially increases with $U$ (Fig.~\ref{fig-2}h),  in agreement with the predicted 
exponentially long lifetime of doublons in the 
Hubbard model for $U\gg \hopp$ \cite{Sensarma2010a}. In the insulator ($U=8$), $\tau$ becomes so large 
that $d(t)$ and $E_\text{kin}(t)$ remain constant on the scale of the plot, apart from a short initial 
transient.

Dissipation completely modifies this picture: Instead of approaching a thermal equilibrium value at 
higher temperature, both $d(t)$ and $E_\text{kin}(t)$ now relax back to their initial values for $U=4$ 
and $U=5$. For $U=4$, this process can be understood within a two-temperature picture
\cite{Allen1987a}:  Because the electronic thermalization time is only a few $\hbar/\hopp$ (Fig.~\ref{fig-2}h), 
electrons can reach a quasi-equilibrium state at high temperature $T^*$ before energy is transferred 
to the environment. The temperature $T^*$ is subsequently reduced towards the fixed temperature of the 
bath with a rate entirely determined by $\lambda$.  This interpretation is supported by the observation 
that the curves $d(t)-d(0)$ for $U=4$ fall on top of each other when the time axis is rescaled by 
$\lambda^2$ (Fig.~\ref{fig-2}g). 
indicating a passage through the same sequence of equilibrium 
states. 
In fact, because $U=4$ is in the metal-insulator crossover region,
the double occupancy behaves non-monotonically as a function of temperature (inset of Fig.~\ref{fig-2}g), 
which explains the minimum in $d(t)$. 

Because the energy quantum that can be transferred to the bath is limited ($\omega_0 = \hopp$), 
coupling to the bosons does not open an efficient channel for the doublon decay for $U \gg \hopp$.
Consequently, the relaxation rate of the double occupancy to its initial value shows a similar 
exponential increase with $U$ as the thermalization time for $\lambda=0$ (Fig.~\ref{fig-2}h),
and the bath has no effect on $d(t)$ in the insulator ($U=8$). The kinetic energy, however, 
does relax to a value that is lower than after the excitation, indicating that doublons and 
holes are cooled to the temperature of the bath before they recombine. In the following 
we will contrast the resulting low-energy state with a chemically doped metal,
by comparing their conductivities.

\begin{figure}[t]
\centerline{ \includegraphics[clip=true,width=0.99\columnwidth]{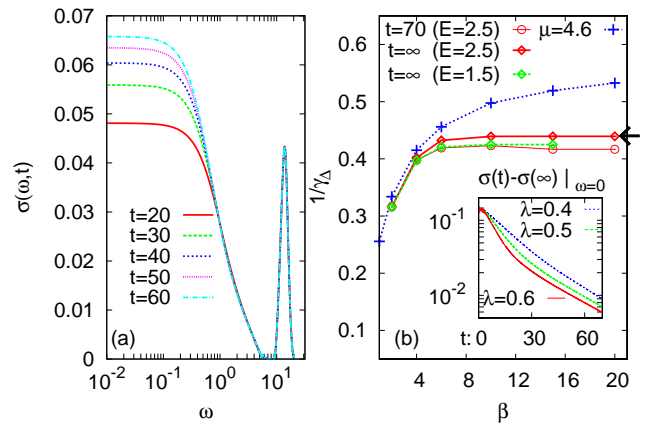}}
\caption{
(a) Optical conductivity  $\sigma(\omega,t)$ [Eq.~(\ref{sigma-av}), $\Delta=0.2$] 
at various times for the photo-doped Mott insulator ($U=14$, $\beta=6$, $\lambda=0.6$; 
Pump: $E_0=2.5$).  (b) Scattering time as a function of (bath) temperature 
for the chemically doped case ($\mu=4.6$) and the photo-doped case. Pump:
$\Omega=14$; excitation density $\delta \approx 0.8 \%$ ($E_0=1.5$) and 
$\delta \approx 2 \%$ ($E_0=2.5$).  Circle symbols: Eq.~(\ref{gamma-heuristic}) taking $\sigma(\omega,t)$ at $t=70$.
Diamonds: Eq.~(\ref{gamma-heuristic}) taking $\sigma(\omega,t)$ from the extrapolation 
to $t=\infty$. Arrow: $\gamma_\Delta$ for a temperature quench ($\delta \approx 2\%$), see text.
(Inset)  $\bar \sigma(\omega,t) - \bar \sigma(\omega,\infty)$ for $\omega=0$,
where $\bar \sigma(\omega,\infty)$ is obtained from a fit $\bar \sigma(\omega,t)
= \bar \sigma(\omega,\infty)+ A \exp(-t/\tau)$.
}
\label{fig-3}
\end{figure}

Figure \ref{fig-3}a shows the optical conductivity of the photo-doped Mott insulator for 
various time-delays after the excitation. The decrease of the kinetic energy 
due to dissipation leads to an increase of spectral weight according to the sum 
rule $\int\!d\omega\,\sigma(\omega,t) = -E_\text{kin}(t)$. Nearly all added weight enters 
the Drude peak, while the weight $W_\text{Hub}$ in the Hubbard band is reduced with 
respect to the undoped case. In fact, 
$W_\text{Hub}$ 
measures only nearest neighbor short-time correlations, and with respect to this, 
photo-doped and chemically-doped states just look alike: the 
reduction of $W_\text{Hub}$ 
is time-independent and proportional to doping $\delta=\Delta d + \Delta h$, where 
here and in the following 
$\Delta d$ and $\Delta h$ 
denote
the differences in the 
doublon and hole densities 
with respect to the un-doped insulator ($\delta \approx 2\%$ in Fig.~\ref{fig-3}). 

The Drude peak and its width $\gamma$, on the other hand, can give a hint on whether or not coherent 
quasi-particles are being formed. Instead of $\gamma$ we compute the approximate measure
\begin{equation}
\label{gamma-heuristic}
\gamma_\Delta 
=
\bar\omega 
\sqrt{
\bar\sigma(\omega=0;\Delta)/\bar\sigma(\bar\omega;\Delta) -1
},
\end{equation}
which can be obtained from a finite time-window
(after relaxation of $E_\text{kin}$), and approaches $\gamma$ for $\Delta\rightarrow 0$ 
and a Lorentz curve (we choose $\bar\omega$ $=$ $\Delta$ $=$ $0.5$).
Although the asymptotic behavior $\gamma\propto T^2$ is cut off below the scale 
$\Delta$,  $\gamma_\Delta$ still  reveals the increase of the inverse scattering rate in the 
interesting temperature range where the chemically doped state enters the Fermi 
liquid regime (Fig.~\ref{fig-3}b). In the photo-doped state, in contrast, $1/\gamma_\Delta$ 
saturates at a smaller value for $T\lesssim 1/5$. Note that $\bar\sigma$ 
is not yet fully stationary at the largest simulation time $t=70$.
However, an exponential extrapolation, $\bar\sigma(t)=\bar\sigma(\infty) + A \exp(-t/\tau)$ 
(Fig.~\ref{fig-3}, inset) corrects $\gamma_\Delta$ only slightly, and moreover, it leads to a 
temperature-independent value (diamond symbols in Fig.~\ref{fig-3}b). 
This finding,  which is a central result of  this paper, is rather insensitive to 
the excitation density (see results for $E_0=1.5$, $\delta \approx 0.8 \%$, and $E_0=2.5$, 
$\delta \approx 1-2 \%$, in Fig.~\ref{fig-3}b). It shows that the photo-doped state does not 
behave like a Fermi-liquid upon lowering the temperature.

\begin{figure}[t]
\centerline{ \includegraphics[clip=true,width=0.99\columnwidth]{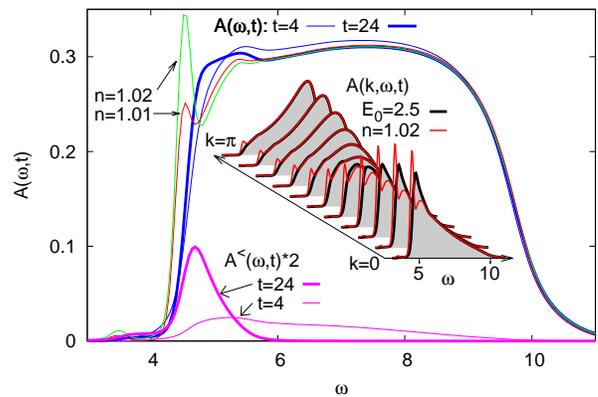}}
\caption{
Spectral function in the upper Hubbard band. 
$A(\omega,t)$, $A^<(\omega,t)$: photo-doped, $U=14$, $\beta=15$, $\lambda=0.6$; pump 
$\Omega=14$, $E_0=2.5$ (excitation density $\Delta d=0.01$). Curves $n=1.01=1+\Delta d$ and $n=1.02=1+2\Delta d$ 
correspond to chemically doped states, where frequency axis is offset by $\Delta\omega=\mu$.
Inset: Momentum-resolved  $A_k(\omega)$ for various $k$ and same parameters. 
Light red curves: chemically doped case at $n=1.02$; shaded curves: photo-doped case. }
\label{fig-4}
\end{figure}

Further differences between photo-doped and chemically doped states are evident from their 
spectral functions (Fig.~\ref{fig-4}): The occupied density of states $A^<(\omega,t) = (1/\pi)\text{Re} \int_0^\infty ds 
e^{i\omega s} \langle  c^\dagger(t+s) c(t) \rangle$, which for a quasi-steady state is related to a 
time-resolved photoemission spectrum  \cite{Freericks2009a}, evolves from a broad photo-induced distribution 
for small times into a peak concentrated at low frequencies. At the same time, the density of 
states, $A(\omega,t) =  (1/\pi)\text{Re} \int_0^\infty ds e^{i\omega s}  \langle  \{c(t+s)^\dagger, c(t)\}
\rangle$, is not rigid as it would be for a noninteracting system, but it develops a feature at 
the lower edge of the upper Hubbard band (the lower Hubbard band is symmetric). 
This feature is different from the quasi-particle peak of a chemically doped system,
both when the density $n$ of the latter is adjusted to the total number of photo-doped 
carriers ($n-1=2\Delta d$) or to the number of electron-like carriers ($n-1=\Delta d$). 
A closer look at the momentum resolved  spectral function $A_k(\omega,t) = (1/\pi)\text{Re} \int_0^\infty ds e^{i\omega s}  
\langle  \{c_k(t+s)^\dagger,c_k(t)\}\rangle$ after the system has become nearly stationary ($t=40$)
reveals that this feature belongs to a relatively broad band of heavily scattered charge excitations (Fig.~\ref{fig-4}, inset).

For $U\gg \hopp$, strong-coupling perturbation theory for the Hubbard model gives a possible
interpretation of our results: Up to corrections of higher order in $\hopp/U$, a Schrieffer-Wolff transformation 
brings the Hamiltonian (\ref{hubbard}) into a form $H_{U\gg \hopp}$ which 
separately conserves both doublon and hole numbers   \cite{Harris1967}. 
Projecting $H_{U\gg \hopp}$ onto the low-energy subspace without doublons (for the hole-doped case) 
or without holes (for the electron-doped case), leads to the familiar $t$-$J$-model. 
In nonequilibrium, however, interesting dynamics can occur in 
different sectors of the Hilbert space, such as metastable superconductivity in the absence of 
singly-occupied states \cite{Rosch2008}, or the strong-coupling prethermalization after an interaction 
quench \cite{Eckstein2009a}. The simplest possible photo-doped state after relaxation is a 
low-temperature state of  $H_{U\gg \hopp}$ with equal density of both doublons and holes.  
Our results hence suggest that this corresponding equilibrium state is a bad metal rather than a 
Fermi liquid in the investigated temperature range, in contrast to the case of pure electron or hole 
doping. At present we cannot directly compute equilibrium properties of $H_{U\gg v}$ for both 
electron and hole doping. However, we can verify that the photo-induced state is rather 
independent of the initial preparation, which is a prerequisite for being a thermal state: We have 
performed a similar numerical analysis as above, but now the system is initially prepared at high 
temperature ($\beta^*=0.4$), and then suddenly coupled to a bath at lower temperature ($\beta=10$). 
In this somewhat artificial temperature-quench, carriers in the system are of thermal origin rather 
than photo-excited, but the long-time behavior of $\gamma_\Delta$ nevertheless turns out to be 
the same as for the photo-doped case (arrow in Fig.~\ref{fig-3}b).

In conclusion, we have demonstrated that the properties of photo-doped and chemically doped Mott 
insulators are quite distinct, even if photo-induced carriers are allowed to transfer their initially high 
kinetic energy to the environment before recombination occurs: A temperature-independent scattering 
time of the carriers in the photo-doped system indicates that when a Mott insulator is doped with 
electrons and holes at the same time, its properties remain non-Fermi liquid like down to much 
lower temperatures  than for pure electron or hole doping.  Experimentally, this results in a low 
mobility of photo-doped carries, which could be seen in a THz-study of the Drude part of photo-doped 
metals. 
On a more fundamental level, our work raises questions concerning the possible ground states of the 
limiting strong-coupling model for the Hubbard model with both electron and hole doping. Does Fermi 
liquid behavior appear below a new, strongly reduced, coherence scale? Should we expect new quantum 
phases formed from paired doublons and holes? Or will this low-energy state simply phase-separate 
into purely hole-doped and electron-doped regions?

We thank J.~Freericks, S.~Kehrein, M.~Kollar,  Th.~Prusch\-ke, T.~Oka, and N.~Tsuji for useful discussions. 
We acknowledge support from the Swiss National Science Foundation (Grant PP002-118866) and 
FP7/ERC starting grant No. 278023.

 \end{document}